\documentclass[12pt,a4paper]{article}

\usepackage{graphicx} 
\usepackage{hyperref}
\usepackage{graphicx}
\usepackage{amssymb}
\usepackage{amsmath}
\usepackage{color}
\usepackage{url}
\usepackage{orcidlink}

\title{\large\bf Dynamical Baryogenesis in Rainbow Cosmology}

\author{\normalsize\sc Surendra Kumar Gour\thanks{\tt s.gour@iitg.ac.in} \ and Malay K. Nandy\,\orcidlink{0000-0002-2640-0497}\thanks{\tt mknandy@iitg.ac.in \rm (Corresponding Author)}\\ 
 {\small\em Department of Physics, Indian Institute of Technology Guwahati, Guwahati 781 039, India}}
 
\date{\small (24 July 2026)} 

\begin{document}

\maketitle

\begin{abstract}
We investigate baryogenesis in the framework of rainbow cosmology employing a complex scalar field with a softly broken global $U(1)$-symmetry. The modified dispersion relation of rainbow cosmology leads to energy-dependent modification to the FLRW metric components, that modifies the Friedmann equation and the scalar field dynamics. We thereby obtain analytical solutions for the scalar field evolution in the radiation-dominated epoch, and show that baryon asymmetry is generated {\em dynamically} even from an initially baryon-symmetric state. We find that the baryon-to-photon ratio asymptotically approaches a constant value in the long-time limit, which is proportional to the symmetry-breaking coupling strength $\lambda$, and scales with the scalar field mass as $M^{-5/2}$. Our results demonstrate that requiring consistency with the observed baryon asymmetry constrains the $\lambda$ and $M$ values within reasonable ranges, thus providing a viable dynamical mechanism for baryogenesis during the radiation-dominated era without invoking supersymmetry.\\

\end{abstract}

\tableofcontents


\section{Introduction}
The observed dominance of baryons over antibaryons in the present universe is one of the most important unresolved problems in cosmology. The Standard Model of particle physics predicts that baryons and antibaryons should have been produced in exactly equal numbers during the early stages of cosmic evolution so that all of them would annihilate into photons in later stages. However, the observed baryons in the universe with almost no antibaryons suggests that baryons must have been produced in excess of antibaryons in the early stages of cosmic evolution.

This baryon asymmetry has been measured with high precision through BBN prediction and CMB WMAP observations \cite{Burles-PRD-2001, Burles-2001, Bennett-2003}, together with the measured baryon density, $\Omega_bh^2 = 0.0224 \pm 0.0001$, reported by the Planck Collaboration \cite{Planck-2018, Planck-2024, PDG-2024}. This asymmetry has been quantified through the baryon-to-photon ratio
\begin{equation}
\eta = \frac{n_b - n_{\bar{b}}}{n_\gamma},
\end{equation}
where $n_b$ and $n_{\bar{b}}$ denote baryon and antibaryon number densities, and $n_{\gamma}$ is the photon density of the universe. Current cosmological measurements indicate that  $\eta$ falls within the interval $(5.5\text{--}6.8)\times10^{-10}$, with a mean value of about $6.1\times10^{-10}$. This is equivalent to the creation of one excess of baryon for every 10 billions of baryon-antibaryon pairs.

A theoretical framework for baryon asymmetry was first proposed by Sakharov \cite{Sakharov-1967}, who identified three necessary conditions for baryogenesis. Various mechanisms for baryogenesis have been formulated to satisfy these conditions, such as GUT baryogenesis \cite{Georgi-Glashow-1974,  Dimopoulos-1978, Yoshimura-1978-A, Yoshimura-1978-B, Weinberg-1979, Kolb-1980, Kolb-1990-book, Weinberg-book}, electro-weak baryogenesis \cite{Kuzmin-1985, Shaposhnikov-1986, Cline-2006}, and leptogenesis \cite{Fukugita-1986}. Another mechanism due to Affleck and Dine \cite{Affleck-Dine-1985, Allahverdi-2012} operates in supersymmetric scenarios.

Moreover, various alternative mechanisms for baryogenesis have been suggested.  Spontaneous baryogenesis \cite{Cohen-1987, Cohen-1988, Cohen-1991} offers a way to generate baryon asymmetry while maintaining thermal equilibrium conditions without requiring $C$ and $CP$ violation. Gravitational baryogenesis \cite{Davoudiasl-2004} explains the origin of baryon asymmetry through the coupling of the baryon current to spacetime curvature whose evolution drives the generation of  baryon asymmetry. This scenario  provides a bridge between cosmological dynamics and fundamental particle interactions.

Rainbow Gravity \cite{Magueijo-2002, Magueijo-2004, Pablo-2004}, a different framework, generalizes the notion of spacetime by allowing the metric to depend on the energy of the probe particle. In this framework,  particles with different energies perceive different spacetime geometries, leading to a rainbow of metrics. This energy dependence modifies the standard cosmological evolution, particularly in the ultra-high energy regime of the early universe.

In such a scenario, the dynamics of the early universe, including the Hubble expansion rate, thermodynamic evolution, dynamical evolution of various degrees of freedom, are altered from their standard cosmological descriptions. These modifications can have important consequences for baryogenesis. In particular, modified dispersion relations and energy-dependent spacetime structure potentially alters baryon production mechanisms, leading to observable deviations in the baryon-to-photon ratio.

It has been shown that gravitational baryogenesis in the framework of $ F(R) $ rainbow gravity can successfully generate the observed baryon asymmetry for appropriate model parameters \cite{Goodarzi-2025}.
The results suggest that quantum gravity corrections encoded through rainbow functions plays an important role in the origin of the baryon-antibaryon asymmetry of the universe.

Consequently, studying baryogenesis within the framework of rainbow gravity provides a promising avenue to connect quantum gravity phenomenology with early-universe cosmology and to probe Planck- and GUT-scale physics through cosmological observations.

In this paper, we therefore investigate analytically the early stages of cosmic evolution within rainbow gravity during the radiation-dominated epoch. In the rainbow framework, we consider a complex scalar field model in which the real and imaginary components represent baryons and antibaryons, respectively, with a softly broken $U(1)$ symmetry. By solving the modified Friedmann equation together with the dynamical equations governing the two field components, we derive the associated Noether charge density. Our analytical calculations indicate that legitimate parameter values can successfully generate the observed baryon asymmetry in the long-time limit.

The remainder of this paper is organized as follows. In Section~\ref{sec-framework}, we present the framework of Rainbow Cosmology. Section~\ref{sec-scalar model} introduces the scalar field model for baryogenesis, while Section~\ref{sec-rainbow model} discusses the choice of the rainbow functions employed in our analysis. In Section~\ref{sec-MFE}, we derive the modified Friedmann equation in the rainbow cosmological framework. Section~\ref{sec-scalar mode} is devoted to the scalar field equations in rainbow cosmology, with the early-time and late-time behaviors analyzed separately. In Section~\ref{sec-photon}, we investigate the photon density in the presence of modified dispersion relation (MDR). Subsequently, derivation of the Noether charge density is carried out in Section~\ref{sec-noether}. Finally, Section~\ref{sec-disc} presents the discussion and conclusion. For completeness, determination of the integration constants is relegated to Appendix~\ref{sec-APP}.

\section{Framework of Rainbow Cosmology} 
\label{sec-framework}

In special relativity, the energy–momentum dispersion relation is given by
\begin{equation}
    E^2 -{\bf p}^2 = m^2,
\label{eq-dr}    
\end{equation}
where $m$ is the rest mass, $E$ is the total energy, and ${\bf p}$ is the momentum of the particle. 

Various models of quantum gravity, such as, doubly special relativity \cite{Camelia-2001, Camelia-2002, Camelia-2002-A}, $\kappa$-deformed noncommutative geometry \cite{Iochum-2011, Arzano-2021}, Liouville string theory \cite{Ellis-1992}, suggest that the dispersion relation is modified at energy scales near the Planck scale. In fact, many models of modified dispersion relation, commonly known as MDR \cite{Magueijo-2002, Magueijo-2003, Camelia-1998, Jacob-2010, Camelia-2008}, have been suggested in the literature. It has also been suggested that such dispersion relations should be important in explaining energy dependent delay in arrival of photons from distant gamma-ray bursts, the delay being due to quantum gravity effects on the intervening vacuum \cite{Camelia-1998}. 

In the context of gravity, a modified dispersion relation can alter the Einstein field equations. Such a theory of gravity is commonly known as rainbow gravity \cite{Magueijo-2002, Magueijo-2004, Pablo-2004}. In the problem of cosmological baryogenesis, the dispersion relation is generally modeled as 
\begin{equation}
     f^{2}(\varepsilon)\,E^2-g^{2}(\varepsilon)\, {\bf p}^2 =m^2,
\label{eq-mdr}     
\end{equation}
where quantum gravity effects near the Planck scale are represented by the deformation functions  $f(\varepsilon)$ and $g(\varepsilon)$, with $\varepsilon=\frac{E}{E_P}$, and $E_P$ is the Planck energy scale. 

To explore the implications of this modified dispersion relation in a cosmological setting, a homogeneous and isotropic expanding universe, with the flat Friedmann-Lemaître-Robertson-Walker (FLRW) metric \cite{Friedmann-1922, Lema-1927, Robertson-1936, Walker-1937, Barrow-2012},
\begin{equation}
    ds^2 = -dt^2 + a^2(t)\,\delta_{ij}\,dx^i dx^j,
\label{eq-flrw}    
\end{equation}
is modified to an effective energy-dependent (rainbow) metric of the form \cite{Magueijo-2002, Magueijo-2004, Pablo-2004}
\begin{equation}
    ds^2 = -\frac{dt^2}{f^{2}(\varepsilon)} + \frac{a^2(t)}{g^{2}(\varepsilon)}\,\delta_{ij}\,dx^i dx^j,
\label{eq-mflrw}    
\end{equation}
where $a(t)$ is the scale factor and $t$ is the cosmic time.

In the low energy limit $E\ll E_P$, or equivalently $\varepsilon\ll 1$, the Rainbow functions must satisfy the conditions, $f(\varepsilon) \to 1 $ and $  g(\varepsilon) \to 1$, so that the modified  dispersion relation \ref{eq-mdr} reduces to the low energy dispersion relation \ref{eq-dr}, and the modified metric \ref{eq-mflrw} reduces to the FLRW metric \ref{eq-flrw}.

\section{Scalar Field Model for Baryogenesis}
\label{sec-scalar model}
 
We model the phenomenon of cosmological baryogenesis with a complex scalar field $\Phi$, having the action 
\begin{equation}
     \mathcal{S}=\int d^{4} x \sqrt{-g}\left[-\frac{1}{2} g^{\mu\nu} \partial_{\mu} \Phi \, \partial_{\nu} \Phi^* -V(\Phi,\Phi^*) \right],
\label{eq-action}     
\end{equation}
where $V(\Phi,\Phi^*)$ is the self-interaction potential. 

The action \ref{eq-action} accounts for the impact of the modified dispersion relation, or equivalently the modified line-element \ref{eq-mflrw}, through the metric $g_{\mu \nu}$. 

Assuming homogeneity and isotropy, the scalar field $\Phi$ may be regarded as a function of time $t$ alone, $\Phi(t)$. Consequently, the action takes the form
\begin{equation}
    \mathcal{S} = V_0\int dt \,\frac{a^3(t)}{f\,g^{3}} \left[\frac{1}{2} f^{2}\, \dot{|\Phi|}^2 -V(\Phi, \Phi^{*}) \right],
\label{eq-action2}    
\end{equation}
where $V_0$ is a fiducial volume. The kinetic part of the action is invariant under global $U(1)$ transformation $\Phi \rightarrow e^{i\alpha}\Phi$. 

Writing $\Phi(t) = u(t) + i\, v(t)$, the action \ref{eq-action2} gives the Lagrangian
\begin{equation}
L = \frac{V_0\,a^3(t)}{f\,g^{3}} \Big[\frac{1}{2}f^2(\dot{u}^2+\dot{v}^2) -V(u,v) \Big],
\label{eq-lag}
\end{equation}
which, through the Euler-Lagrange equation, yields the equations of motion as two second-order diﬀerential equations,
\begin{equation}
\ddot{u} + \left(3H + \frac{\dot{f}}{f} - 3\frac{\dot{g}}{g} \right)\dot{u} + \frac{1}{f^{2}} \frac{\partial V}{\partial u} = 0
\label{eq-u}
\end{equation}
and
\begin{equation}
\ddot{v} + \left(3H + \frac{\dot{f}}{f} - 3\frac{\dot{g}}{g} \right)\dot{v} + \frac{1}{f^{2}} \frac{\partial V}{\partial v} = 0,
\label{eq-v}
\end{equation}
where $ H = \frac{\dot{a}}{a}$ is the Hubble parameter. 

Planck-scale corrections to the last terms enter through the rainbow factor $\frac{1}{f^{2}}$ rescaling the effective potential gradient, while Hubble friction is modified by acquiring corrections through logrithimic derivatives of the rainbow functions $f(\varepsilon)$ and $g(\varepsilon)$. 

In the low-energy limit $f(\varepsilon) \to 1$ and $g(\varepsilon) \to 1$, these equations \ref{eq-u} and \ref{eq-v} reduce to 
\begin{equation}
\ddot{u} + 3H \dot{u} + \frac{\partial V}{\partial u} = 0 \quad {\rm and} \quad \ddot{v} + 3H \dot{v} + \frac{\partial V}{\partial v}  = 0.
\end{equation}
in FLRW spacetime. 

In this work, we model the potential as
\begin{equation}
V\left(\Phi, \Phi^{*}\right)= \frac{1}{2} M^{2}\left|\Phi\right|^{2}- \frac{\lambda}{4} M^{2}\left(\Phi^{2}+\Phi^{*2}\right),
\label{eq-pot}
\end{equation}
where the second part of the potential, containing the coupling constant $\lambda$, softly breaks the global $U(1)$-symmetry (for $\lambda \ll 1$), that is, under the transformation $\Phi \to e^{i\alpha} \Phi$.

This potential \ref{eq-pot} can be expressed as a function of real and imaginary components of $\Phi= u+iv$ as
\begin{equation}
V(u,v) =\frac{1}{2}M_u^{2}\,u^2 +\frac{1}{2}M_v^{2}\,v^2,
\label{eq-V}
\end{equation}
where $M_u^2= M^2(1-\lambda)$ and $M_v^2=M^2(1+\lambda)$.

Thus, the $U(1)$-breaking part of the potential \ref{eq-pot} generates a very small mass splitting between the two scalar components $u(t)$ and $v(t)$, since $\lambda \ll 1$.

\section{Model of Rainbow Functions}
\label{sec-rainbow model}

We model the rainbow functions as
\begin{equation}
f(\varepsilon) = 1 + \frac{E}{2E_P} \quad \text{and} \quad g(\varepsilon) = 1,
\label{eq-rf}
\end{equation}
motivated by the work of Amelino-Camelia et al.~\cite{Camelia-1998}.

Using the rainbow functions \ref{eq-rf} in \ref{eq-u} and \ref{eq-v}, the equations of motion assume the forms
\begin{equation}
\ddot{u} + \left[3H + \frac{\dot{E}}{E + 2 E_P}\right]\dot{u} + \frac{1}{(1 + \frac{E}{2E_P})^2} \frac{\partial V}{\partial u} = 0
\label{eq-u2}
\end{equation}
and
\begin{equation}
\ddot{v} + \left[3H + \frac{\dot{E}}{E + 2 E_P}\right]\dot{v} + \frac{1}{(1 + \frac{E}{2E_P})^2} \frac{\partial V}{\partial v} = 0,
\label{eq-v2}
\end{equation}
and the dispersion relation \ref{eq-mdr} becomes
\begin{equation}
\left(1+\frac{E}{2E_P}\right)^2 E^2 - {\bf p}^2 = m^2.
\label{eq-mdr2}
\end{equation}
Differentiating \ref{eq-mdr2} with respect to the cosmic time $t$, we get
\begin{equation}
\dot E=\frac{2\,{\bf p}\cdot\dot {\bf p}}{2E+\dfrac{3E^2}{E_P}+\dfrac{E^3}{E_P^2}}.
\label{eq-Edot}
\end{equation}

In the expanding universe, the physical momentum undergoes red shift, so that 
\begin{equation}
\dot {\bf p} = -H{\bf p},
\end{equation}
and equation \ref{eq-Edot} takes the form
\begin{equation}
\dot E=-\frac{2H{\bf p}^2}{2E+\dfrac{3E^2}{E_P}+\dfrac{E^3}{E_P^2}}.
\label{eq-Edot2}
\end{equation}
Eliminating ${\bf p}^2$ using \ref{eq-mdr2}, the right-hand side of the energy evolution equation~\ref{eq-Edot2} can be written entirely in terms of $E$, and we have
\begin{equation}
\dot E=-\frac{2H\left[\left(1+\dfrac{E}{2E_P}\right)^2E^2-m^2\right]}{2E+\dfrac{3E^2}{E_P}+\dfrac{E^3}{E_P^2}}.
\label{eq-Ed}
\end{equation}

For the massless case ($ m=0 $), and in the low-energy limit $E\ll E_P$, we can approximate \ref{eq-Ed} to
\begin{equation}
\dot E = -HE\left(1-\frac{E}{2E_P}\right),
\label{eq-Edot3}
\end{equation}
in the first leading order correction, so that equations \ref{eq-u2} and \ref{eq-v2} reduce to
\begin{equation}
\ddot{u}+\left(3H-\frac{HE}{2E_P}\right)\dot{u}+ \frac{1}{\left(1+\dfrac{E}{E_P}\right)}\frac{\partial V}{\partial u}= 0
\label{eq-u3}
\end{equation}
and
\begin{equation}
\ddot{v}+\left(3H-\frac{HE}{2E_P}\right)\dot{v}+ \frac{1}{\left(1+\dfrac{E}{E_P}\right)}
\frac{\partial V}{\partial v} = 0.
\label{eq-v3}
\end{equation}

Noting that $H=\frac{\dot{a}}{a}$, equation \ref{eq-Edot3} can be rewritten as 
\begin{equation}
a\frac{dE}{da}= -E\left(1-\frac{E}{2E_P}\right).
\label{eq-dada}
\end{equation}

Separating variables, and integrating equation \ref{eq-dada}, we obtain $E$ as a function of the scale factor $a$,
\begin{equation}
    \frac{E}{2E_P}= \frac{K}{K+a},
\label{eq-Ea}    
\end{equation}
where $K$ is the integration constant. Since $E>0$ and $a>0$, we must have $K>0$. ($K$ cannot be negative as $E$ would then blow up for $a=K$.)

\section{Modified Friedmann Equation}
\label{sec-MFE}

Using a generic rainbow metric \ref{eq-mflrw}, the Einstein field equation $R_{\mu\nu}-\frac{1}{2}Rg_{\mu\nu}=8 \pi G\, T_{\mu\nu}$ leads to a modified Friedmann equation \cite{Magueijo-2004, Ling-2007},
\begin{equation}
H^2= \frac{8\pi G}{3}\frac{\rho_r}{f^{2}},
\label{eq-mfe}
\end{equation}
where $\rho_r$ is the energy density of the radiation-dominated universe in this regime, where the radiation energy density dominates the cosmic energy density. 

Since $\rho_r \propto a^{-4}$ for radiation, we have
\begin{equation}
\rho_r=\rho_{0}\,a^{-4},
\label{eq-rd}
\end{equation}
where $\rho_{0}$ is a constant.

Noting that $H=\frac{\dot{a}}{a}$ and substituting equation \ref{eq-rd} together with the rainbow functions \ref{eq-rf}, equation \ref{eq-mfe} gives the evolution for the scale-factor,
\begin{equation}
\dot a=\frac{\sqrt{\dfrac{8\pi G}{3}\rho_{0}}}{1+\dfrac{E}{2E_P}}a^{-1}.
\label{eq-adot}
\end{equation}

Using equation \ref{eq-Ea} in \ref{eq-adot}, and noting that $\left(1+\dfrac{E}{2E_P}\right)^{-1} \simeq \left(1-\dfrac{E}{2E_P}\right)$ for $E \ll E_P$, we obtain
\begin{equation}
\dot a=\frac{\sqrt{\frac{8\pi G}{3}\rho_{0}}}{K+a}.
\label{eq-aK}
\end{equation}
Integrating equation \ref{eq-aK}, we obtain
\begin{equation}
\frac{a^2}{2}+Ka=\sqrt{\frac{8\pi G}{3}\rho_{0}}\,t + K_1.
\label{eq-aK1}
\end{equation}
Since the scale factor $a$ becomes very small at the initial time $t\to 0$, we take the integration constant  $K_1=0$. Consequently, equation \ref{eq-aK1} yields
\begin{equation}
a(t)=-K\pm\sqrt{ K^2 +\Gamma t },
\label{eq-aK2}
\end{equation}
where 
\begin{equation}
\Gamma=\sqrt{\frac{32\pi G\rho_0}{3}}.
\end{equation}
Noting that $a>0$ and $K>0$ (that follows from equation \ref{eq-Ea}), we must choose the positive sign for the discriminant part of equation \ref{eq-aK2}, leading to
\begin{equation}
a(t)=-K+\sqrt{ K^2 +\Gamma t }.
\label{eq-a}
\end{equation}

We can employ equation \ref{eq-a} to calculate the Hubble parameter as a function of the cosmic time $t$, yielding 
\begin{equation}
    H(t) = \frac{\dot a}{a}  = \frac{1}{2t}\left(1+\frac{K}{\sqrt{K^2+ \Gamma t}}\right).
\label{eq-HC}    
\end{equation}

Using equation \ref{eq-a}, the Hubble parameter can also be expressed as a function of the  scale factor $a$, 
\begin{equation}
H(a)=\frac{\Gamma}{2a(K+a)}.
\label{eq-Ha}
\end{equation}

For sufficiently small time $t$, we have $\frac{1}{\sqrt{K^2+\Gamma t}} = \frac{1}{K}\left(1 - \frac{\Gamma t}{2K^2} + \frac{3 \Gamma^2 t^2}{8K^4} \cdots \right)$, so that the early-time behavior of the Hubble parameter \ref{eq-HC} is
\begin{equation}
H(t)=\frac{1}{t} - \frac{\Gamma}{4K^2} + \frac{3 \Gamma^2 t}{16K^4} -  \ldots.
\end{equation}
Considering the leading order behaviour $H(t)= \frac{1}{t}$, the scale factor is obtained as
\begin{equation}
a(t) \sim t.
\end{equation}
This linear scaling is due to a pronounced effect of the rainbow function at early times.

For the late time behaviour, we have $\frac{1}{\sqrt{K^2+\Gamma t}}= \frac{1}{\sqrt{\Gamma t}}\left(1 - \frac{K^2}{2 \Gamma t} + \frac{3K^4}{8 \Gamma^2 t^2} \cdots \right)$, and therefore equation \ref{eq-HC} yields
\begin{equation}
H(t)=\frac{1}{2t}+\frac{K}{2\sqrt{\Gamma}}\,t^{-3/2}-\frac{K^3}{4 \Gamma^{3/2}}\,t^{-5/2}+ \ldots.
\end{equation}
Upon considering the leading order behaviour $H(t)= \frac{1}{2t}$, the scale factor is obtained as
\begin{equation}
a(t) \sim t^{1/2}.
\end{equation}
This reduction to the normal scaling for radiation suggests that the effect of the rainbow function is considerably diminished at late times.

\section{Scalar Field Dynamics in Rainbow Cosmology}
\label{sec-scalar mode}

Using equations \ref{eq-Ha} and \ref{eq-Ea}, the dynamical equations \ref{eq-u3} and \ref{eq-v3} assume the forms 
\begin{equation}
\ddot u + \gamma(a)\dot u + \beta(a)\frac{\partial V}{\partial u}=0
\label{eq-u4}
\end{equation}
and
\begin{equation}
\ddot v + \gamma(a)\dot v + \beta(a) \frac{\partial V}{\partial v}=0,
\label{eq-v4}
\end{equation}
where 
\begin{equation}
\gamma(a)= \frac{\Gamma (2K+3a)}{2a(K+a)^2} \quad {\rm and} \quad \beta(a)=\frac{K+a}{3K+a}.
\end{equation}
Employing equation \ref{eq-a}, $\gamma(a)$ and $\beta(a)$ can be expressed as functions of time~$t$,
\begin{equation}
\gamma(t)=\frac{\Gamma \left(3\sqrt{K^2+ \Gamma t}-K\right)}{2(\sqrt{K^2+ \Gamma t}-K)(K^2+ \Gamma t)} \quad {\rm and} \quad \beta(t) = \frac{\sqrt{K^2+ \Gamma t}}{\sqrt{K^2+ \Gamma t}+2K}.
\label{eq-gb}
\end{equation}

Upon using equation \ref{eq-V}, the field equations \ref{eq-u4} and \ref{eq-v4} can be written in the form
\begin{equation}
\ddot u+\gamma(t)\dot u+\omega_u^2(t)u=0
\label{eq-master1}
\end{equation}
and
\begin{equation}
\ddot v+\gamma(t)\dot v+\omega_v^2(t)v=0,
\label{eq-master2}
\end{equation}
where $\omega_u^2(t)= M_u^2 \, \beta(t)$ and $\omega_v^2(t)= M_v^2 \, \beta(t)$.

\subsection{Early-Time Behavior}

For small times, we have $\sqrt{K^2+ \Gamma t}=K+\frac{\Gamma t}{2K}-\frac{\Gamma^2 t^2}{8K^3}+O(t^3)$, so that from \ref{eq-gb}, the early time behaviour reads
\begin{equation}
\gamma(t) = \frac{2}{t}\left[1-\frac{\Gamma t}{2K^2}-\frac{11 \Gamma^2 t^2}{16K^4}+O(t^3)\right]
\label{eq-ge}
\end{equation}
and
\begin{equation}
\beta(t) = \frac{1}{3}\left[1+\frac{\Gamma t}{3K^2}-\frac{11\Gamma^2 t^2}{36K^4}+O(t^3)\right].
\label{eq-be}
\end{equation}
 
Taking the leading order contributions to $\gamma(t)$ and  $\beta(t)$ from \ref{eq-ge} and \ref{eq-be}, the field equations \ref{eq-u4} and \ref{eq-v4} reduce to
\begin{equation}
\ddot u+\frac{2}{t}\dot u+\omega_u^2 u=0 \quad {\rm and} \quad \ddot v+\frac{2}{t}\dot v+\omega_v^2 v=0,
\end{equation}
where $\omega_u=\frac{M_u}{\sqrt{3}} $ and $  \omega_v=\frac{M_v}{\sqrt{3}}$.

Substituting $u(t)=\frac{\chi_u(t)}{t}$ and $v(t)=\frac{\chi_v(t)}{t}$, these evolution equations transform to  
\begin{equation}
\ddot \chi_u+\omega_u^2 \chi_u=0 \quad {\rm and} \quad \ddot \chi_v+\omega_v^2 \chi_v=0.
\end{equation}
These equations have general solutions 
\begin{equation}
\chi_u(t)=A_u\sin(\omega_u t)+B_u\cos(\omega_u t)
\end{equation}
and
\begin{equation}
\chi_v(t)=A_v\sin(\omega_v t)+B_v\cos(\omega_v t).
\end{equation}
Thus,
\begin{equation}
u(t)=\frac{1}{t}\left[A_u\sin\,\left(\frac{M_u}{\sqrt{3}}t\right) +B_u\cos\,\left(\frac{M_u}{\sqrt{3}}t\right)\right]
\end{equation}
and
\begin{equation}
v(t)=\frac{1}{t}\left[A_v\sin\,\left(\frac{M_v}{\sqrt{3}}t\right)+B_v\cos\,\left(\frac{M_v}{\sqrt{3}}t\right)\right].
\end{equation}
Since $u(t)$ and $v(t)$ should not diverge in the limit $t\to0$, we must set $B_u=B_v=0$. Consequently, physically acceptable solutions are 
\begin{equation}
u(t)=\frac{A_u}{t}\sin\,\left(\frac{M_u}{\sqrt{3}}t\right) \quad {\rm and} \quad v(t)=\frac{A_v}{t}\sin\,\left(\frac{M_v}{\sqrt{3}}t\right).
\label{eq-uvsol}
\end{equation}

These are early-time solutions obtained with $\gamma \approx \frac{2}{t}$ and $\beta \approx \frac{1}{3}$ that are fairly good approximations valid for {\em small times} as evident from equations \ref{eq-ge} and \ref{eq-be}.

\subsection{Late-Time Behavior}

For long times, we have $\sqrt{K^2+\Gamma t}=\sqrt{\Gamma t}+\frac{K^2}{2\sqrt{\Gamma t}}-\frac{K^4}{8(\Gamma t)^{3/2}}+ O(t^{-5/2})$, so that from \ref{eq-gb}, the late time behaviour reads
\begin{equation}
\gamma(t)=\frac{3}{2t}\left[1+\frac{2K}{3\sqrt{\Gamma t}}+\frac{2K^2}{3\Gamma t} + O(t^{-3/2})\right]
\label{eq-gl}
\end{equation}
and
\begin{equation}
\beta(t)=\left[1-\frac{2K}{\sqrt{\Gamma t}}+\frac{4K^2}{\Gamma t}-O(t^{-3/2})\right].
\label{eq-bl}
\end{equation}

Taking the leading order contributions to $\gamma(t)$ and  $\beta(t)$ from \ref{eq-gl} and \ref{eq-bl}, the field equations \ref{eq-u4} and \ref{eq-v4} reduce to
\begin{equation}
\ddot u+\frac{3}{2t}\dot u+M_u^2 u=0 \quad {\rm and} \quad \ddot v+\frac{3}{2t}\dot v+M_v^2 v=0,
\label{eq-uv1}
\end{equation}
in the late-time regime.

Substituting $u(t)= t^{-1/4}\chi_u(t)$ and $v(t)= t^{-1/4}\chi_v(t)$, these evolution equations \ref{eq-uv1} transform to  
\begin{equation}
t^2\ddot\chi_u+t\dot\chi_u+\left(M_u^2t^2-\frac{1}{16}\right)\chi_u=0
\end{equation} 
and
\begin{equation}
t^2\ddot\chi_v+t\dot\chi_v+\left(M_v^2t^2-\frac{1}{16}\right)\chi_v=0,
\end{equation}
which are Bessel equations of order $\nu=\frac14$. These equations have general solutions 
\begin{equation}
\chi_u(t)=C_u J_{1/4}(M_u t)+D_u Y_{1/4}(M_u t)
\end{equation}
and
\begin{equation}
\chi_v(t)=C_v J_{1/4}(M_v t)+D_v Y_{1/4}(M_v t),
\end{equation}
where $J_{\nu}(x)$ and  $Y_{\nu}(x)$ are Bessel functions of the first and second kind of order $\nu=1/4$.

Thus, the original field solutions assume the forms
\begin{equation}
u(t)=t^{-1/4}\left[C_u J_{1/4}(M_u t)+D_u Y_{1/4}(M_u t)\right]
\label{eq-usol}
\end{equation}
and 
\begin{equation}
v(t)=t^{-1/4}\left[C_v J_{1/4}(M_v t)+D_v Y_{1/4}(M_v t)\right].
\label{eq-vsol}
\end{equation}

These are late-time solutions obtained with approximations  $\gamma \approx \frac{3}{2t}$ and $\beta \approx 1$ valid for {\em large times} as seen from equations \ref{eq-gl} and \ref{eq-bl}. Moreover, both Bessel functions are physically acceptable since they are well-behaved in the regime $t>0$.

Noting that the Bessel functions admit the asymptotic forms \cite{Arfken-1995, Abramowitz-1964}
\begin{align}
J_\nu(x) \approx \sqrt{\frac{2}{\pi x}} \cos\left(x - \frac{\nu\pi}{2} - \frac{\pi}{4}\right), \label{eq-J}\\
Y_\nu(x) \approx \sqrt{\frac{2}{\pi x}} \sin\left(x - \frac{\nu\pi}{2} - \frac{\pi}{4}\right),
\label{eq-Y}
\end{align}
for infinitely large $x \to \infty $, equations \ref{eq-usol} and \ref{eq-vsol} assume the asymptotic forms 
\begin{equation}
u(t) \approx t^{-1/4}\sqrt{\frac{2}{\pi M_u t}}\left[C_u \cos\left(M_u t - \frac{3\pi}{8}\right)+ D_u \sin\left(M_u t - \frac{3\pi}{8}\right)\right]
\end{equation}
and 
\begin{equation}
v(t) \approx t^{-1/4}\sqrt{\frac{2}{\pi M_v t}}\left[C_v \cos\left(M_v t - \frac{3\pi}{8}\right)+ D_v \sin\left(M_v t - \frac{3\pi}{8}\right)\right],
\end{equation}
 in the long-time limit $t \to \infty $.
 
These forms indicate that the solutions assume damped oscillatory behaviour at very long times.

\section{Photon Density with MDR}
\label{sec-photon}

With the modified dispersion relation (MDR) \ref{eq-mdr}, momentum and energy are related by ${\bf p}^2= \frac{f^2(\varepsilon)}{g^2(\varepsilon)}\,E^2 $ for photons ($m=0$). Using this relation, photon number density defined by
\begin{equation}
    n_\gamma = 2 \int \frac{d^3{\bf p}}{(2\pi)^3} \frac{1}{e^{E/T}-1},
\end{equation}
takes the general form
\begin{equation}
n_\gamma=\frac{1}{\pi^2}\int_0^\infty \dfrac{f^2}{g^2} \left[\dfrac{f}{g}+E\dfrac{d}{dE}\left(\dfrac{f}{g}\right)\right] \frac{E^2dE}{e^{E/T}-1}.
\label{eq-ngam}
\end{equation}

Substituting the rainbow functions \ref{eq-rf} into \ref{eq-ngam}, we obtain
\begin{equation}
    n_\gamma=\frac{1}{\pi^2}\int_0^\infty\left(1+\dfrac{E}{2E_P}\right)^2\left(1+\dfrac{E}{E_P}\right) \frac{E^2dE}{e^{E/T}-1}.
\label{eq-ngam0}    
\end{equation}
Retaining correction to first-order in $\frac{E}{E_P}$, equation \ref{eq-ngam0} reduces to
\begin{equation}
    n_\gamma=\frac{1}{\pi^2}\left[\int_0^\infty\frac{E^2\, dE}{e^{E/T}-1}+\frac{2}{E_P}\int_0^\infty
    \frac{E^3\, dE}{e^{E/T}-1}\right].
\end{equation}
Carrying out the integrations, we finally arrive at
\begin{equation}
n_\gamma= \frac{2\zeta(3)}{\pi^2}T^3+\frac{2\pi^2}{15}\frac{T^4}{E_P},
\label{eq-ngam2}
\end{equation}
which represents the photon-number density due to the modified dispersion relation. Here $\zeta(3)=1.2021$ is the Riemann zeta function $\zeta(z)$ for $z=3$.

The ratio of the second term to the first term in equation~\ref{eq-ngam2} is $\frac{\pi^4}{15\zeta(3)}\frac{T}{E_P}=5.4\,\frac{T}{E_P}$. Thus, the second term adds a negligible amount of correction for the legitimate assumption $T \sim E_{\rm GUT} \ll E_P$.

Therefore, this rainbow gravity correction term, and all higher order corrections in powers of $\frac{T}{E_P}$, are highly suppressed, and the standard radiation term, that is the first term in equation \ref{eq-ngam2}, dominates. 

Using the modified Friedmann equation \ref{eq-mfe} with radiation energy density $\rho_r = \frac{\pi^2}{30}g_{*}\,T^4+\mathcal{O}(\frac{T^5}{E_P})$, and employing the rainbow function \ref{eq-rf}, we obtain the temperature as a function of time $t$,
\begin{equation}
T(t) = \left(\frac{45}{4\pi^3 G g_*}\right)^{1/4} \sqrt{H(t)\left(1 + \frac{E}{2E_P}\right)}.
\label{eq-Tt}
\end{equation}
Using equations \ref{eq-HC}, \ref{eq-Ea}, and \ref{eq-a}, equation \ref{eq-Tt} yields
\begin{equation}
T(t) = \left(\frac{45}{8\pi^3 G g_*}\right)^{1/4} \left(1 + \frac{K}{\sqrt{K^2+\Gamma t}}\right)\,t^{-1/2}.
\label{eq-Tt2}
\end{equation}

Substituting equation \ref{eq-Tt2} into the expression for photon number density~\ref{eq-ngam2}, and assuming $T\ll E_P$ leads to
\begin{equation}
n_\gamma(t)=\frac{2\zeta(3)}{\pi^2}\left(\frac{45}{8\pi^3 G g_*}\right)^{3/4}\left(1 + \frac{K}{\sqrt{K^2+\Gamma t}}\right)^3t^{-3/2}.
\end{equation}

In the long-time limit $t\to \infty$, this expression reduces to
\begin{equation}
n_\gamma(t)\simeq\frac{2\zeta(3)}{\pi^2}\left(\frac{45}{\pi^2 g_*}\right)^{3/4} M_P^{3/2}\,t^{-3/2},
\label{eq-pd1}
\end{equation}
in the leading order contribution, where $M_P=\frac{1}{\sqrt{8 \pi G}}$ is the reduced Planck mass.

\section{Noether Charge Density}
\label{sec-noether}

Using Noether theorem, the Lagrangian \ref{eq-lag} gives the $U(1)$ charge density as
\begin{equation}
 \rho= \frac{i}{2}\frac{f}{g^{3}}\left(\Phi \dot{\Phi}^{*}-\Phi^{*}\dot{\Phi}\right) = \frac{f}{g^{3}}\left(u\dot{v}-v\dot{u}\right).
 \label{eq-ncd0}
\end{equation}
Employing the rainbow functions \ref{eq-rf}, together with \ref{eq-Ea} and \ref{eq-a}, equation \ref{eq-ncd0} yields 
\begin{equation}
\rho(t)=\frac{a+2K}{a+K}\left(u\dot{v}-v\dot{u}\right) =\left(1+\frac{K}{\sqrt{K^2+ \Gamma t}}\right)\left(u\dot{v}-v\dot{u}\right).
\label{eq-ncd}
\end{equation}

For the early time behaviour, using equations \ref{eq-uvsol} in \ref{eq-ncd}, we get
\begin{equation}
\rho(t)= \frac{2A_uA_v}{\sqrt{3}\,t^2}\left[M_v\sin\left(\frac{M_u t}{\sqrt{3}}\right)\cos\left(\frac{M_v t}{\sqrt{3}}\right)-M_u\sin\left(\frac{M_v t}{\sqrt{3}}\right)\cos\left(\frac{M_u t}{\sqrt{3}}\right)\right]
\end{equation}
in the leading order, which further reduces to
\begin{equation}
\rho(t)=\frac{2}{27}A_uA_v\,M_uM_v\,(M_u^2-M_v^2)\,t+O(t^3),
\end{equation}
so that $\rho \to 0$ at very early times, $t\to 0$. On the other hand, $\dot{\rho} \to$ constant since 
\begin{equation}
\dot{\rho}(t)=\frac{2}{27}A_uA_v\,M_uM_v\,(M_u^2-M_v^2)+O(t^2),
\end{equation}

For the late time behaviour, using equations \ref{eq-usol} and \ref{eq-vsol} in \ref{eq-ncd}, we get
\begin{align}
\rho(t)=t^{-1/2}\Big[&M_v\Big\{C_uJ_{1/4}(M_ut)+D_uY_{1/4}(M_ut)\Big\}\Big\{C_vJ'_{1/4}(M_vt)+D_vY'_{1/4}(M_vt)\Big\}\nonumber\\[2mm]&-M_u\Big\{C_vJ_{1/4}(M_vt)+D_vY_{1/4}(M_vt)\Big\}\Big\{C_uJ'_{1/4}(M_ut)+D_uY'_{1/4}(M_ut)\Big\}\Big]
\end{align}
in the leading order. 

Using the asymptotic forms \ref{eq-J} and \ref{eq-Y} for the Bessel functions, these solutions further reduce to
\begin{align}
\rho(t) \approx&\frac{2}{\pi\sqrt{M_uM_v}}\,t^{-3/2}\Bigg[M_v\Bigg\{C_u\cos\,\left(M_u t-\frac{3\pi}{8}\right)
+D_u\sin\,\left(M_u t-\frac{3\pi}{8}\right)\Bigg\}\nonumber\\&\qquad\times\Bigg\{-C_v\sin\,\left(M_v t-\frac{3\pi}{8}\right)+D_v\cos\,\left(M_v t-\frac{3\pi}{8}\right)\Bigg\}\nonumber\\&\qquad-M_u\Bigg\{C_v\cos\,\left(M_v t-\frac{3\pi}{8}\right)+D_v\sin\,\left(M_v t-\frac{3\pi}{8}\right)\Bigg\}\nonumber\\&\qquad\times\Bigg\{-C_u\sin\,\left(M_u t-\frac{3\pi}{8}\right)+D_u\cos\,\left(M_u t-\frac{3\pi}{8}\right)\Bigg\}\Bigg],
\label{eq-nc}
\end{align}
in the long time limit, $t \to \infty$.

Since the coupling constant $\lambda \ll 1 $, we can apply the approximations $M_u^2=M^2(1-\lambda) \approx M^2$ and $M_v^2=M^2(1+\lambda) \approx M^2$, leading to
\begin{equation}
\begin{array}{c}
\sin\,\left(M_u t-\frac{3\pi}{8}\right) \approx \sin\,\left(M_v t-\frac{3\pi}{8}\right)  \approx \sin\,\left(M t-\frac{3\pi}{8}\right),\\
\cos\,\left(M_u t-\frac{3\pi}{8}\right)  \approx \cos\,\left(M_v t-\frac{3\pi}{8}\right)  \approx \cos\,\left(M t-\frac{3\pi}{8}\right).
\end{array}
\label{eq-approx}
\end{equation}

Using the above approximations \ref{eq-approx}, we obtain from equation \ref{eq-nc},
\begin{align}
\rho(t)\approx& \frac{1}{\pi\sqrt{M_uM_v}}\,t^{-3/2}\Big[(M_u+M_v)(C_uD_v-C_vD_u)\nonumber \\
+&(M_v-M_u)(C_uD_v+C_vD_u)\cos\Big(2Mt-\frac{3\pi}{4}\Big)\nonumber \\
+&(M_v-M_u)(D_uD_v-C_uC_v)\sin\Big(2Mt-\frac{3\pi}{4}\Big)\Big].
\label{eq-rhoav1}
\end{align}

Taking time average over many periods of oscillation, we have
\begin{equation}
\left\langle
\cos\left(2Mt-\frac{3\pi}{4}\right)\right\rangle=0,\quad \left\langle \sin\left(2Mt-\frac{3\pi}{4}\right) \right\rangle=0,
\end{equation}
so that only the non-oscillatory term in equation \ref{eq-rhoav1} contributes. Therefore, the time averaged Noether charge density becomes
\begin{equation}
\left\langle \rho(t)\right\rangle=\frac{M_u+M_v}{\pi\sqrt{M_uM_v}}(C_uD_v-C_vD_u)\,t^{-3/2},
\label{eq-rhoav2}
\end{equation}
in the long time limit, $t \to \infty$.

Since $(C_uD_v-C_vD_u)=\mathcal O(\lambda)$ as shown in Appendix \ref{sec-APP}, we have to pick up zeroth order contributions from the rest of the factors in \ref{eq-rhoav2}. Noting that
\begin{equation}
M_u+M_v=2M+\mathcal{O}(\lambda^{2}) \quad \text{and} \quad \sqrt{M_uM_v}=M+\mathcal{O}(\lambda^{2}), 
\end{equation}
we thus have
\begin{equation}
\left\langle \rho(t)\right\rangle=\frac{2}{\pi}(C_uD_v-C_vD_u)\,t^{-3/2}+\mathcal{O}(\lambda^3).
\label{eq-avrho}
\end{equation}

Thus, the time-averaged Noether charge density exhibits the power-law decay $t^{-3/2}$.

The coefficients $C_u, D_u$ and $C_v, D_v$ are calculated in Appendix \ref{sec-APP} by matching the early and late time solutions at the crossover time $t_c \approx \frac{1}{M}$, that is inverse of the scalar mass. It is shown in equation \ref{eq-diff2} of Appendix \ref{sec-APP} that
\begin{equation}
C_uD_v-C_vD_u= \frac{3\pi}{8}\,\lambda\,A_uA_v\,M^{3/2}\sin^2\left(\frac{1}{\sqrt3}\right)+\mathcal{O}(\lambda^2).
\label{eq-cd1}
\end{equation}

This expression \ref{eq-cd1} shows that baryogenesis can procesed even if the two components $u(t)$ and $v(t)$ start with the same amplitude in the initial phase, that is, $A_u=A_v= A$.

Using equation \ref{eq-cd1} in equation \ref{eq-avrho}, we thus obtain the average Noether charge density in the form
\begin{equation}
\left\langle \rho(t)\right\rangle=\frac{3}{4}\,\lambda\,A^{2}\,M^{3/2}\sin^2\left(\frac{1}{\sqrt3}\right)\,t^{-3/2},
\label{eq-avrho1}
\end{equation}
in the long time limit, $t \to \infty$.

The kinetic part of the Lagrangian \ref{eq-lag} vanishes at the initial time $t=0$ since $\dot{u}(0)=\dot{v}(0)=0$, which can be seen from the early time solution given by equations \ref{eq-uvsol}. Consequently, the energy of the scalar field resides entirely in the potential part initially at $t=0$.

\begin{table}[h!]
\centering
\caption{\small \sf Values of the coupling constant $\lambda$ and the corresponding values of the scalar mass $M$ (as calculated from equation \ref{eq-etafinal}) constrained by the observed mean value of baryon asymmetry $\eta$.}
\begin{tabular}{cccc}
\hline
$\eta$ & $\lambda$ & $M~(\mathrm{GeV})$ & Mass Scale \\
\hline
$6.1\times10^{-10}$ &$10^{-7}$    & $1.59\times10^{16}$ & $ \ E_{\rm GUT}$\\
$6.1\times10^{-10}$ &$10^{-10}$   & $1.00\times10^{15}$ & $10^{-1} \ E_{\rm GUT}$ \\
$6.1\times10^{-10}$ &$10^{-12}$   & $1.59\times10^{14}$ &  $10^{-2} \ E_{\rm GUT}$\\
$6.1\times10^{-10}$ &$10^{-15}$   & $1.00\times10^{13}$ &  $10^{-3} \ E_{\rm GUT}$\\
$6.1\times10^{-10}$ &$10^{-18}$   & $1.59\times10^{12}$ &  $10^{-4} \ E_{\rm GUT}$\\
\hline
\end{tabular}
\label{table1}
\end{table}

Assuming that baryogenesis initiates at the GUT scale $E_{\rm GUT} \sim 10^{-2} \,M_P$, the initial potential energy density, as given by the Lagrangian \ref{eq-lag}, is obtained as  
\begin{equation}
 U_0=\left.\frac{1}{fg^{3}}\, V(u,v)\right|_{t=0}= E^{4}_{\rm GUT}.
 \label{eq-U0}
\end{equation}
Employing equations \ref{eq-rf}, \ref{eq-V}, and \ref{eq-uvsol} in \ref{eq-U0}, we thus get 
\begin{equation}
 A = \sqrt{3}\, \left(\frac{E_{\rm GUT}}{M}\right)^{2} \left(1+ \frac{E_{\rm GUT}}{2E_P}\right)^{1/2} +\mathcal{O}(\lambda^2),
 \label{eq-A}
\end{equation}
with a baryon symmetric state $A_u=A_v=A$ at $t=0$.

From equations  \ref{eq-avrho1} and \ref{eq-pd1}, along with \ref{eq-A}, we obtain  
\begin{equation}
\eta=\frac{\left\langle \rho(t)\right\rangle}{n_\gamma} \approx \frac{9 \pi^2}{8 \zeta(3)}\, \left(\frac{\pi^2 g_{*}}{45}\right)^{3/4} \lambda\, \left(\frac{\rm E_{GUT}}{M}\right)^{4}\, \left(\frac{M}{M_P}\right)^{3/2} \sin^2\left(\frac{1}{\sqrt3}\right),
\end{equation}
in the long time limit, $t \to \infty$.

With $g_{*}=106.75$, the eﬀective number of relativistic degrees of freedom in the Standard Model of particle physics, we finally obtain the baryon-to-photon ratio as 
\begin{equation}
\eta=\frac{\left\langle \rho(t)\right\rangle}{n_\gamma} \approx 29.2872777\, \lambda\, \left(\frac{E_{\rm GUT}}{M}\right)^{4}\, \left(\frac{M}{M_P}\right)^{3/2},
\label{eq-etafinal}
\end{equation}
where $E_{\rm GUT}=10^{16}$ GeV \cite{Amaldi-1991, Cheng-1997, Bourilkov-2015} and  $M_P=2.435 \times 10^{18}$ GeV.

We show in Table \ref{table1} different values of the coupling constant $\lambda$ and the corresponding values of the scalar mass $M$ constrained by the observed mean value of baryon asymmetry, $\eta = 6.1\times10^{-10}$.

\section{Discussion and Conclusion}
\label{sec-disc}

In this work, we investigated a dynamical mechanism for baryogenesis within the framework of rainbow cosmology using a model of complex scalar field with a potential softly breaking the global $U(1)$ symmetry, thereby inducing a small mass splitting between the two components of the field. The modified dispersion relation characterized by the rainbow framework introduces energy-dependent rainbow functions into the FLRW metric, leading to modifications of the Friedmann equation and the scalar field equations of motion.

Within this framework, we considered baryogenesis during the radiation-dominated epoch, where the radiation energy density dominates the cosmic energy density. Analytical solutions to the equations of motion were obtained in the early- and late-time regimes with appropriate approximations. These solutions showed that early-time evolution is described by damped sinusoidal oscillatory solutions, whereas the late-time behaviour is a linear combination of Bessel functions of the first and second kinds. The inclusion of the Bessel function of the second kind is justified because it remains well behaved throughout the physically relevant region $t>0$.

To construct a continuous cosmological evolution, the early- and late-time solutions, together with their first derivatives, were matched at a crossover time in the order of the inverse mass $M^{-1}$ of the scalar field.

An important outcome of the present analysis is that a  baryon asymmetry is generated dynamically even when the initial state is baryon symmetric. With this initially symmetric state, the amplitudes of the early-time solutions become equal and can be related to the initial potential energy of the scalar field, which is naturally identified with the GUT energy scale. 

Upon determining all integration constants through matching the solutions in the two regimes at the crossover time $M^{-1}$, the average Noether charge density was evaluated in the long-time limit. It was found to be proportional to the $U(1)$ symmetry-breaking coupling strength $\lambda$ at leading order, while decaying as $t^{-3/2}$.

The photon number density $n_{\gamma}$ was also derived employing the modified dispersion relation in the rainbow framework. The corrections arising from higher powers of $T/E_P$ are strongly suppressed, so that the standard leading-order dependence on temperature $n_\gamma \propto T^3$ dominates. Furthermore, the modified Friedmann equation reproduces the conventional temperature-time relation $T\propto t^{-1/2}$ in the long-time limit, implying the standard scaling $n_\gamma\propto t^{-3/2}$ in the asymptotic regime, $t \to \infty$.

Combining these results, the baryon-to-photon ratio $\eta$ approaches a constant value  in the asymptotic regime, $t \to \infty$. The resulting asymmetry is proportional to the symmetry-breaking coupling strength $\lambda$ and scales with the mass of scalar field as $M^{-5/2}$. The parameters $\lambda$ and $M$ are thus naturally constrained by requiring consistency with the observed baryon asymmetry. Upon considering the range $10^{-4}E_{\rm GUT}\lesssim M\lesssim E_{\rm GUT}$, the symmetry-breaking coupling is found to be within $10^{-18}\lesssim \lambda \lesssim 10^{-7}$. These values are physically reasonable and demonstrate that the proposed dynamical mechanism can successfully account for the observed baryon asymmetry.

Overall, the present work demonstrates that rainbow cosmology, combined with a softly broken global $U(1)$ symmetry, provides a viable dynamical framework for the  generation of the observed baryon asymmetry during the radiation-dominated epoch. The mechanism naturally accommodates physically reasonable scalar field masses and symmetry-breaking couplings without invoking supersymmetry. These results suggest that Planck-scale modifications of spacetime geometry plays a nontrivial role in the origin of the matter-antimatter asymmetry, motivating further investigations of baryogenesis in quantum gravity inspired cosmological models.

\section*{Acknowledgements}
Surendra Kumar Gour acknowledges the financial assistance received through a Research Fellowship from the Ministry of Education, Government of India.

\appendix

\section{Determination of the Integration Constants}
\label{sec-APP}

We obtained early and late time analytical solutions of the differential equations \ref{eq-master1} and \ref{eq-master2} by assuming that a crossover happens between these two regimes at time $t_c =\frac{K^2}{\Gamma}$ while approximating the expressions \ref{eq-gb}. We also identify this crossover time with the inverse mass scale $t_c =\frac{1}{M}$.

The early time solutions \ref{eq-uvsol} (denoted $u_e$ and $v_e$ below) and the late time solutions \ref{eq-usol} and \ref{eq-vsol} (denoted $u_l$ and $v_l$ below) must be matched at the transition point $t_c$. Consequently, we employ the following matching conditions:
\begin{equation}
u_e(t_c) = u_l(t_c)
\label{eq-uel}
\end{equation}
\begin{equation}
\dot{u}_e(t_c) = \dot{u}_l(t_c)
\label{eq-duel}
\end{equation}
\begin{equation}
v_e(t_c) = v_l(t_c)
\label{eq-vel}
\end{equation}
\begin{equation}
\dot{v}_e(t_c) = \dot{v}_l(t_c)
\label{eq-dvel}
\end{equation}

These conditions yield four relations among the integration constants appearing in the early time solutions \ref{eq-uvsol} and the late time solutions \ref{eq-usol} and \ref{eq-vsol}. 

Continuity of the field component $u(t)$ and its first derivative at the transition time $t_c$ given by \ref{eq-uel} and \ref{eq-duel} yield two matching conditions:
\begin{equation}
A_u t_c^{-3/4}s=C_u J_{1/4}(x)+D_u Y_{1/4}(x)
\label{eq-system1}
\end{equation}
and
\begin{equation}
A_u\left[-\frac{3}{4M_u}t_c^{-7/4}s+\frac{1}{\sqrt{3}}t_c^{-3/4}c\right]= C_u J'_{1/4}(x)+D_u Y'_{1/4}(x),
\label{eq-system2}
\end{equation}
where we have introduced the notation
\begin{equation}
x=M_u t_c, \quad s=\sin\,\left(\frac{x}{\sqrt{3}}\right),\quad c=\cos\,\left(\frac{x}{\sqrt{3}}\right).
\end{equation}

Equations \eqref{eq-system1} and \eqref{eq-system2} can be written in matrix form as
\begin{equation}
\begin{pmatrix}
J_{1/4}(x) & Y_{1/4}(x)\\[2mm]
J'_{1/4}(x) & Y'_{1/4}(x)
\end{pmatrix}
\begin{pmatrix}
C_u\\[2mm]
D_u
\end{pmatrix}
=
A_u
\begin{pmatrix}
t_c^{-3/4}s\\[2mm]
-\dfrac{3}{4M_u}t_c^{-7/4}s
+\dfrac{1}{\sqrt{3}}t_c^{-3/4}c
\end{pmatrix}.
\end{equation}
Using Cramer's rule, we find
\begin{equation}
C_u=\frac{A_u}{W(x)}\left[t_c^{-3/4}s\,Y'_{1/4}(x)-\left(-\frac{3}{4M_u}t_c^{-7/4}s+\frac{1}{\sqrt{3}}t_c^{-3/4}c\right)Y_{1/4}(x)\right]
\label{eq-Cu}
\end{equation}
and
\begin{equation}
D_u=\frac{A_u}{W(x)}\left[\left(-\frac{3}{4M_u}t_c^{-7/4}s+\frac{1}{\sqrt{3}}t_c^{-3/4}c\right)J_{1/4}(x)-t_c^{-3/4}s\,J'_{1/4}(x)\right],
\label{eq-Du}
\end{equation}
where the wronskian is given by 
\begin{equation}
W(x)=J_{1/4}(x)Y'_{1/4}(x)-J'_{1/4}(x)Y_{1/4}(x)=\frac{2}{\pi x},
\end{equation}
the last equality follows from an identity for the Bessel functions \cite{Arfken-1995, Abramowitz-1964}.

Continuity of the field component $v(t)$ and its first derivative at the transition time $t_c$ given by \ref{eq-vel} and \ref{eq-dvel} yield another two matching conditions that can be identically calculated and obtained immediately from the above relations by the following replacements:
\begin{equation}
A_u\to A_v,\quad M_u\to M_v,\quad C_u\to C_v,\quad D_u\to D_v.
\label{eq-replace}
\end{equation}

Equations \ref{eq-Cu} and \ref{eq-Du}, upon returning to the original quantities, assume the following expressions connecting the integration constants appearing in the early and late time solutions of the field component $u(t)$: 
\begin{align}
C_u &= \frac{\pi M_u t_c}{2} A_u \Bigg[t_c^{-3/4}\sin\,\left(\frac{M_u}{\sqrt{3}}t_c\right)\,Y'_{1/4}(M_u t_c)
- \frac{1}{\sqrt{3}} t_c^{-3/4}\cos\,\left(\frac{M_u}{\sqrt{3}}t_c\right)\,Y_{1/4}(M_u t_c) \notag\\&\hspace{4.2em}
+ \frac{3}{4M_u} t_c^{-7/4}\sin\,\left(\frac{M_u}{\sqrt{3}}t_c\right)\,Y_{1/4}(M_u t_c)\Bigg], \label{eq-cu}\\[1.2ex]
D_u &= \frac{\pi M_u t_c}{2} A_u \Bigg[- t_c^{-3/4}\sin\,\left(\frac{M_u}{\sqrt{3}}t_c\right)\,J'_{1/4}(M_u t_c)
+ \frac{1}{\sqrt{3}} t_c^{-3/4}\cos\,\left(\frac{M_u}{\sqrt{3}}t_c\right)\,J_{1/4}(M_u t_c) \notag\\&\hspace{4.2em}
-\frac{3}{4M_u} t_c^{-7/4}\sin\,\left(\frac{M_u}{\sqrt{3}}t_c\right)\,J_{1/4}(M_u t_c)\Bigg].
\label{eq-du}
\end{align}

Similarly, upon employing the replacements \ref{eq-replace} in equations \ref{eq-cu} and \ref{eq-du}, we obtain the following expressions connecting the integration constants appearing in the early and late time solutions of the field component $v(t)$: 
\begin{align}
C_v &= \frac{\pi M_v t_c}{2} A_v \Bigg[t_c^{-3/4}\sin\,\left(\frac{M_v}{\sqrt{3}}t_c\right)\,Y'_{1/4}(M_v t_c)
- \frac{1}{\sqrt{3}} t_c^{-3/4}\cos\,\left(\frac{M_v}{\sqrt{3}}t_c\right)\,Y_{1/4}(M_v t_c) \notag\\&\hspace{4.2em}
+ \frac{3}{4M_v} t_c^{-7/4}\sin\,\left(\frac{M_v}{\sqrt{3}}t_c\right)\,Y_{1/4}(M_v t_c)\Bigg], \label{eq-cv}\\[1.2ex]
D_v &= \frac{\pi M_v t_c}{2} A_v \Bigg[- t_c^{-3/4}\sin\,\left(\frac{M_v}{\sqrt{3}}t_c\right)\,J'_{1/4}(M_v t_c)
+ \frac{1}{\sqrt{3}} t_c^{-3/4}\cos\,\left(\frac{M_v}{\sqrt{3}}t_c\right)\,J_{1/4}(M_v t_c) \notag\\&\hspace{4.2em} 
-\frac{3}{4M_v} t_c^{-7/4}\sin\,\left(\frac{M_v}{\sqrt{3}}t_c\right)\,J_{1/4}(M_v t_c)\Bigg].
\label{eq-dv}
\end{align}

Since the coupling constant $\lambda \ll 1 $, we can apply the approximations $M_u^2=M^2(1-\lambda) \approx M^2$ and $M_v^2=M^2(1+\lambda) \approx M^2$, so that
\begin{equation}
\begin{array}{c}
\sin\left(\frac{M_u}{\sqrt{3}}t_c\right) \approx \sin\left(\frac{M_v}{\sqrt{3}}t_c\right)\ \approx \sin\left(\frac{M}{\sqrt{3}}t_c\right)\ = \sin\left(\frac{1}{\sqrt{3}}\right), \\
\cos\left(\frac{M_u}{\sqrt{3}}t_c\right) \approx \cos\left(\frac{M_v}{\sqrt{3}}t_c\right) \approx \cos\left(\frac{M}{\sqrt{3}}t_c\right) =  \cos\left(\frac{1}{\sqrt{3}}\right), \\
J_{1/4}(M_u t_c) \approx J_{1/4}(M_v t_c) \approx J_{1/4}(M t_c) = J_{1/4}(1), \\
J'_{1/4}(M_u t_c) \approx J'_{1/4}(M_v t_c) \approx J'_{1/4}(M t_c) = J'_{1/4}(1), \\
Y_{1/4}(M_u t_c) \approx Y_{1/4}(M_v t_c) \approx Y_{1/4}(M t_c) = Y_{1/4}(1), \\
Y'_{1/4}(M_u t_c) \approx Y'_{1/4}(M_v t_c) \approx Y'_{1/4}(M t_c)= Y'_{1/4}(1), 
\end{array}
\label{eq-array}
\end{equation}
since the crossover time is identified as $t_c=\frac{1}{M}$. 

Using the approximations \ref{eq-array}, we obtain from equations \ref{eq-cu}, \ref{eq-du}, \ref{eq-cv} and \ref{eq-dv},
\begin{equation}
C_uD_v-C_vD_u=\frac{3\pi^2}{16}A_uA_v\,t_c^{-1/2}\left(M_v-M_u\right) \left[Y'_{1/4}(1)J_{1/4}(1)-Y_{1/4}(1)J'_{1/4}(1)\right] \sin^2\left(\frac{1}{\sqrt3}\right).
\label{eq-diff1}
\end{equation}
Since $(M_v-M_u)=M \lambda$, the $\mathcal O(\lambda)$ corrections to the zeroth order approximations from the rest of the factors in \ref{eq-diff1} contribute insignificantly at $\mathcal O(\lambda^2)$.

Using the Wronskian relation for the Bessel functions \cite{Arfken-1995, Abramowitz-1964}, 
\begin{equation}
J_\nu(x)Y'_\nu(x)-J'_\nu(x)Y_\nu(x)=\frac{2}{\pi x},
\end{equation}
equation \ref{eq-diff1} yields 
\begin{equation}
C_uD_v-C_vD_u= \frac{3\pi}{8}\lambda A_uA_v M^{3/2} \sin^2\left(\frac{1}{\sqrt3}\right) +\mathcal{O}(\lambda^2).
\label{eq-diff2}
\end{equation}
This result \ref{eq-diff2} has been used in equation \ref{eq-avrho} to evaluate the time-averaged Noether charge density.


\end{document}